\documentclass[a4paper,11pt]{article}
\usepackage{pos}
\usepackage{booktabs}
\usepackage{hyperref}

\title{Improving predictions for the $pp \to t\bar{t}W^+$ process at the LHC with the MINLO method}
\ShortTitle{The $pp \to t\bar{t}W^+$ process with the MINLO method}

\author*[a]{Nikolaos Dimitrakopoulos}

\affiliation[a]{Institute for Theoretical Particle Physics and Cosmology, RWTH Aachen University,
D-52056 Aachen, Germany}

\emailAdd{ndimitrak@physik.rwth-aachen.de}

\abstract{We compare the standard NLO and $\texttt{MiNLO}$ approaches for the full off-shell $pp \to t\bar{t}W^+ + X$ process at the LHC, at both the integrated and differential cross-section levels. Contrary to NLO, in the \texttt{MiNLO} approach, which is now also available within the \textsc{Helac-Nlo} framework, renormalization and factorization scales are dynamically extracted and Sudakov form factors are incorporated. Results are particularly presented for the $pp \to t\bar{t}W^+j$ process at NLO accuracy in perturbative QCD using two different scale choices. Merged predictions up to two jets to improve the overall modeling of the full off-shell $pp \to t\bar{t}W^+ + X$ process are also discussed.\\

\begin{flushright} P3H-26-080,  
TTK-26-38 \end{flushright}}

\FullConference{The 33rd International Workshop on Deep Inelastic Scattering and Related Subjects (DIS2026)\\
4 - 8 May 2026\\
Bologna, Italy\\}

\begin{document}
\maketitle

\section{Introduction}
The production of a top-antitop pair in association with an electroweak $W^\pm$ gauge boson is of particular interest at the LHC. At leading order (LO) in perturbative QCD, $t\bar{t}W^\pm$ production proceeds exclusively through $q\bar{q}'$ initial states, while gluon-initiated channels first appear at next-to-leading order (NLO). Moreover, this process is relevant for Standard Model (SM) measurements and searches for new physics, in particular those involving same-sign leptons, 
and provides sensitivity to four-quark interactions.
On the experimental side, $pp \to t\bar{t}W^\pm$ production has been measured by both the CMS and ATLAS collaborations, see Refs. \cite{ATLAS:2024moy,CMS:2022tkv,CMS:2025iwa} for the most recent analyses. Tensions between theory predictions and data persist, with a mild tension of approximately $2\sigma$ at the inclusive level and discrepancies in several differential distributions. Although these tensions do not constitute evidence for new physics, more accurate theoretical predictions are required to draw more precise conclusions.
On the theory side, an NNLO QCD calculation \cite{Becchetti:2026awn} of on-shell $t\bar{t}W^\pm$ production has recently become available, with the two-loop amplitudes computed in the leading-color approximation. Nonetheless, a more realistic description of the fiducial phase space requires accounting for the decay products of the $t\bar{t}W^\pm$ system, for which full off-shell predictions are available \cite{Bevilacqua:2020pzy,Denner:2020hgg,Denner:2021hqi}. Furthermore, the importance of additional jet activity has been demonstrated in a full off-shell calculation of $pp \to t\bar{t}W^+j + X$ \cite{Bi:2023ucp} with NLO QCD accuracy for the additional jet. In this study, the fiducial $t\bar{t}W^+j$ cross section was found to constitute a substantial fraction of the fiducial $t\bar{t}W^+$ cross section, while at the differential level, significant distortions relative to $t\bar{t}W^+$ production were identified. Since these results are sensitive to the renormalization $(\mu_R$) and factorization $(\mu_F)$ scales chosen in the presence of an additional light jet, it is desirable to compare the standard NLO approach with alternative methods that aim to reduce the ambiguity associated with the scale choice. One such approach is the \texttt{MiNLO} method, introduced in Ref. \cite{Hamilton:2012np}, and built upon the CKKW procedure \cite{Catani:2001cc,Krauss:2002up}. In this framework, $\mu_R$ and $\mu_F$ are dynamically determined from the event kinematics by identifying a probable branching history through the application of an inverse-$k_T$ clustering algorithm \cite{Ellis:1993tq,Catani:1993hr}. In addition, Sudakov form factors are included to resum large logarithms that arise when a large hierarchy of scales is present.

In this work, we summarize some of the results of Ref. \cite{Dimitrakopoulos:2026jwi}, where a comparison between NLO and \texttt{MiNLO} was presented for the full off-shell $pp \to t\bar{t}W^+ + X$ process. In Sec. \ref{sec:minlo_implementation}, we briefly review \texttt{MiNLO}, highlighting the differences from our implementation, and describe our computational setup. In Sec. \ref{sec:nlo_vs_minlo}, we compare the standard NLO and \texttt{MiNLO} predictions for $pp \to t\bar{t}W^+j + X$ at the integrated and differential cross-section levels, while Sec. \ref{sec:merging} discusses the merging of $pp \to t\bar{t}W^+ + \mathrm{jets}$ samples with different jet multiplicities. Finally, we summarize our findings in Sec. \ref{sec:conclusions}.

\section{\texttt{MiNLO} implementation and computational setup} \label{sec:minlo_implementation}
In Ref. \cite{Dimitrakopoulos:2026jwi}, the \texttt{MiNLO} method, which has now been implemented in the \textsc{Helac-Nlo} framework \cite{Bevilacqua:2011xh}, was applied for the first time to a process capturing full off-shell effects. For a detailed description of the method, the reader is referred to the original implementation \cite{Hamilton:2012np} as well as Refs. \cite{Moretti:2016jnv, Dimitrakopoulos:2026jwi}. Here, we briefly summarize the method and highlight the main differences between Ref. \cite{Hamilton:2012np} and our implementation in \textsc{Helac-Nlo} that are relevant for this study. To reduce the ambiguity associated with the scale choice, \texttt{MiNLO} identifies a core system of particles that define the hard process, while the scales associated with additional jet emissions are dynamically extracted using an inverse-$k_T$ clustering algorithm after identifying the most likely sequence of branchings. In our study, the core system comprises the decay products of the $t\bar{t}W^+$ system, namely the $pp \to e^+ \nu_e \mu^- \bar{\nu}_\mu \tau^+ \nu_\tau b\bar{b}$ process. The scale associated with the core system is referred to as the core scale, $q_{core}$, while the scales associated with the reconstructed branchings are given by the corresponding $k_T$ measures and denoted as nodal scales, $q_k$. These scales enter as arguments of the additional $\alpha_s$ couplings with respect to those from the core process, while at NLO, an appropriate subtraction is performed to avoid double counting and to maintain the NLO accuracy of the computation. Once the event skeleton has been constructed, Sudakov form factors are applied to all external and internal lines. The main differences between the original implementation and our implementation in \textsc{Helac-Nlo} concern the treatment of the Sudakov form factors. In particular, unlike the original implementation, we require probabilistic Sudakov form factors that cannot exceed unity. The exact formulas for these Sudakov form factors are provided in Refs. \cite{Hoche:2016elu,Moretti:2016jnv,Dimitrakopoulos:2026jwi}. Furthermore, under scale variations, closely following Ref. \cite{Hoche:2016elu}, the Sudakov form factors associated with the incoming lines are varied consistently to match the PDF evolution up to $\mu_F$. 

In what follows, we compare the standard NLO QCD calculation for the full off-shell $pp \to t\bar{t}W^+j + X$ process in the multilepton channel, as presented in Ref. \cite{Bi:2023ucp}, with results obtained using the \texttt{MiNLO} method. In the former approach, $\mu_R$ and $\mu_F$ are user-defined, while in the latter, the only arbitrary scale is $q_{core}$, with the remaining nodal scales determined dynamically by the \texttt{MiNLO} procedure. We present results using both a dynamical $\mu_0 = E_T/2$ and a fixed $\mu_0 = m_t + m_W/2$ scale choice. The dynamical scale is defined as
\begin{equation} \label{eq:ETscale}
E_T = \sqrt{m_t^2 + p_{T,t}^2}
+ \sqrt{m_t^2 + p_{T,\bar{t}}^2}
+ \sqrt{m_W^2 + p_{T,W}^2} + p_{T,j_1}\,,
\end{equation}
where the kinematics of the $t$, $\bar{t}$, and $W^+$ boson are determined by minimizing
\begin{equation}
Q = |m_t - M_t| + |m_t - M_{\bar{t}}| + |m_W - M_W|\,,
\end{equation}
where $M_t$, $M_{\bar{t}}$, and $M_W$ are the reconstructed invariant masses of the top quark, antitop quark, and $W^+$ boson, respectively, obtained by considering all possible $b,\ell,\nu_\ell$ assignments. It should be noted that, in the \texttt{MiNLO} procedure, the above scale definitions correspond to the $q_{core}$ scale. Since the additional light jet is not part of the core system, its transverse momentum is generally absent from the definition of Eq. \eqref{eq:ETscale}. Furthermore, following Ref. \cite{Hamilton:2012np}, in the case of an unordered clustering, i.e. $q_1 > q_{core}$, the core scale is forced to be equal to $q_1$. Finally, to estimate the size of the theory uncertainties associated with missing higher-order corrections, we employ the standard seven-point scale variation. In the \texttt{MiNLO} approach, the exact scale variation procedure follows the original implementation of Ref. \cite{Hamilton:2012np}, with the exception of varying the scale in the Sudakov form factors of the incoming lines under $\mu_F$ variation, as previously mentioned.

\section{Comparison between NLO and \texttt{MiNLO} for the full off-shell $pp \to t\bar{t}W^+\,j + X$ process} \label{sec:nlo_vs_minlo}
In Table \ref{tab:ttwj}, we compare the standard NLO and \texttt{MiNLO} methods for the full off-shell $pp \to t\bar{t}W^+ j + X$ process in the multilepton channel at the LHC with $\sqrt{s} = 13 \, \rm TeV$, using the $\mu_0 = E_T/2$ and $\mu_0 = m_t + m_W/2$ scale choices.
\begin{table}[!t]
\centering
\scalebox{1.0}{
\begin{tabular}{cccc}
\midrule\midrule
$\mu_0$  & $\sigma^{\textrm{NLO}}$ [ab]  &  $\sigma^{\texttt{MiNLO}}$ [ab] & $\dfrac{\sigma^\texttt{MiNLO}}{\sigma^{\rm NLO}}$\\

\midrule\midrule
 $m_t + m_W/2$ & $144.7^{\,\,\,+0\%}_{-14\%}$  &  $140.9^{\,\,\,+2\%}_{-11\%}$ & 0.97\\
\midrule
 $E_T/2$ & $140.1^{\,\,\,+4\%}_{-10\%}$  &  $136.5^{\,\,\,+6\%}_{-10\%}$ & 0.97\\
\midrule
\end{tabular}}
\caption{\textit{Integrated fiducial cross sections for the full off-shell $pp \to t\bar{t}W^+ j + X$ process in the multilepton channel at the LHC with $\sqrt{s} = 13 \, \rm TeV$, obtained using the standard NLO and \texttt{MiNLO} approaches, using the $\mu_0 = E_T/2$ defined in Eq. \eqref{eq:ETscale} and $\mu_0 = m_t + m_W/2$ scale choices. Also shown are scale uncertainties from the standard seven-point scale variation as well as the ratio of \texttt{MiNLO} to NLO in the last column.}}
\label{tab:ttwj} 
\end{table}
First, we find that the differences between the two approaches are of the order of $3\%$ only and are fully covered by the scale uncertainties. The latter are found to be smaller upon applying the \texttt{MiNLO} method with the fixed scale choice, with the reduction being from $14\%$ to $11\%$. However, when well-behaved scale choices are employed, such as the $\mu_0 = E_T/2$ scale in this case, \texttt{MiNLO} and standard NLO are found to behave similarly, yielding theory uncertainties of the same size.
\begin{figure}[b!]
        \centering        
        \includegraphics[width=0.48\linewidth]{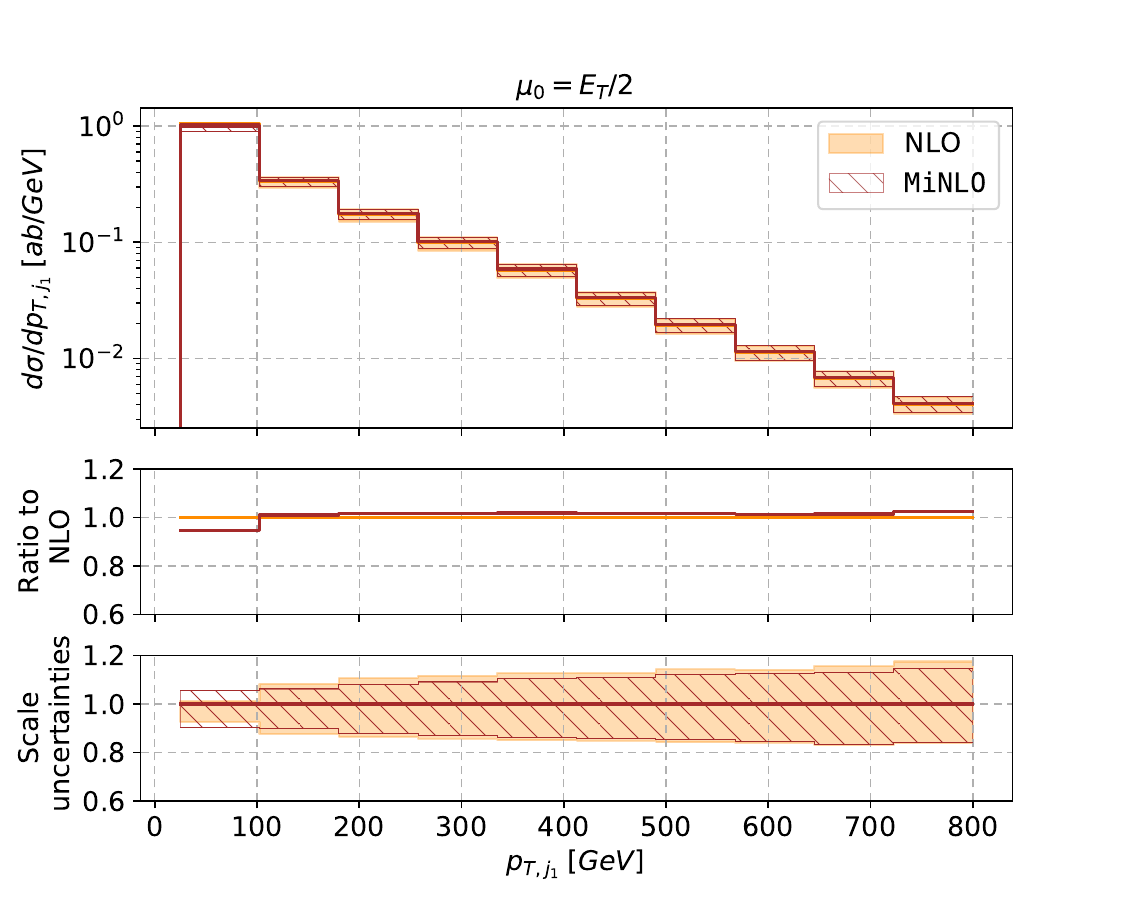}
        \includegraphics[width=0.48\linewidth]{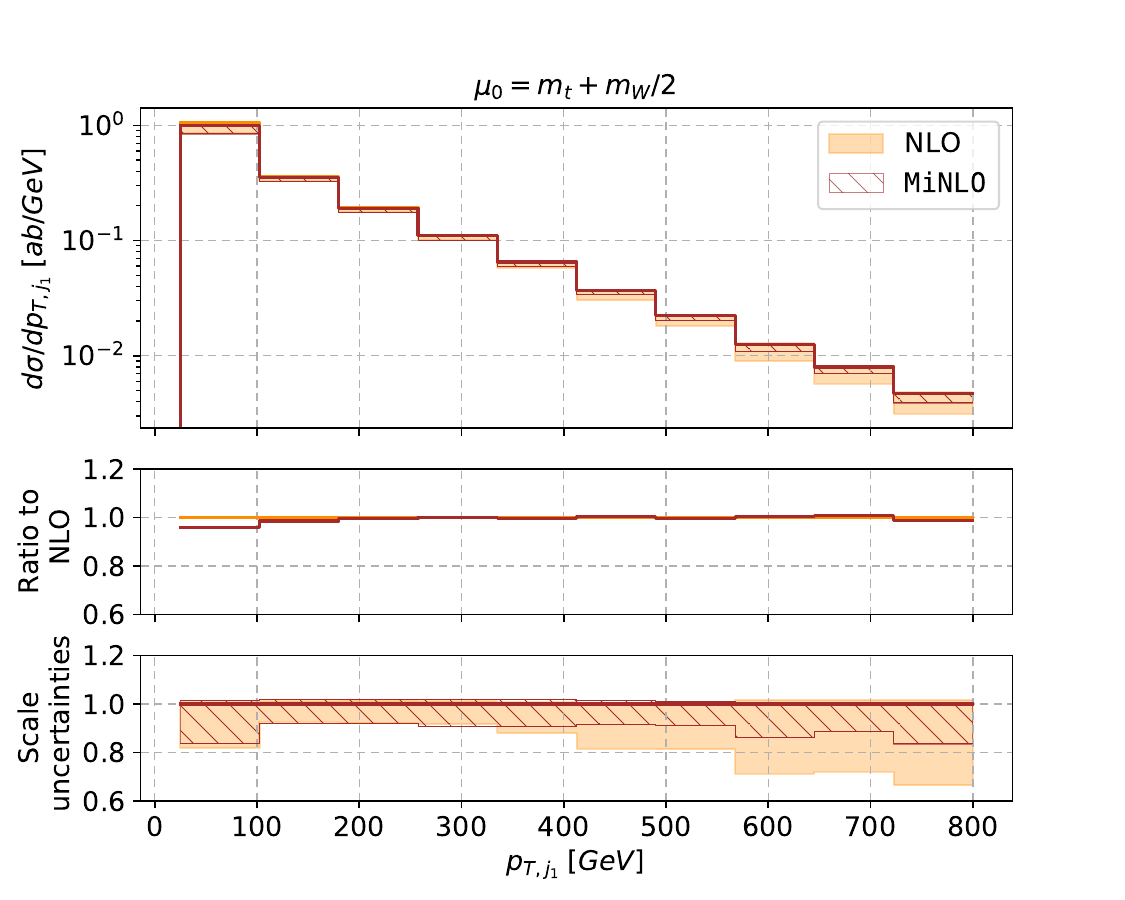}
        \caption{\textit{Differential cross-section comparison between NLO and \texttt{MiNLO} for the full off-shell $pp \rightarrow t\bar{t}W^+\,j + X $ process at the LHC with $\sqrt{s} = 13 \, \rm TeV$, for the transverse momentum of the hardest light jet $p_{T,\,j_1}$. Results are shown using a dynamical $\mu_0 = E_T/2$ (left) and a fixed $\mu_0 = m_t + m_W/2$ (right) scale choice. The upper panels display the absolute predictions, the middle panels the ratio to NLO, and the bottom panels the corresponding scale uncertainties.}}
         \label{fig:minlovsnlo_ptj}
\end{figure}
A differential comparison between NLO and \texttt{MiNLO} is also depicted in Fig. \ref{fig:minlovsnlo_ptj} using the $\mu_0 = E_T/2$ and $\mu_0 = m_t + m_W/2$ scale choices for the $p_{T,\,j_1}$ observable. Similar to the results obtained at the integrated level, for the dynamical scale choice, the two approaches yield comparable results with only minor deviations, while the theory uncertainties are of the same size. However, when the fixed-scale choice is employed, \texttt{MiNLO} performs better in the tail of the distribution, with uncertainty bands becoming almost half as large as those from standard NLO. This is due to the increased probability of an unordered clustering, $(q_1 > q_{core})$, in this regime, which forces $q_{core} = q_1$ and thus removes the problematic scale choice from the calculation. Hence, the large NLO uncertainties in the tails of certain dimensionful observables obtained with a fixed scale are reduced upon applying the \texttt{MiNLO} method. It should be highlighted, however, that other dimensionful observables that do not involve light jets in their definition, such as leptonic observables, are less affected by \texttt{MiNLO}, and potentially large NLO uncertainties in the tails of their distributions remain. 
This is expected, since the \texttt{MiNLO} procedure primarily affects the dynamics of light jets, which are directly involved in the inverse-$k_T$ clustering algorithm.

\section{Multi-jet merged predictions} \label{sec:merging}
\begin{figure}[b!]
        \centering        
        \includegraphics[width=0.48\linewidth]{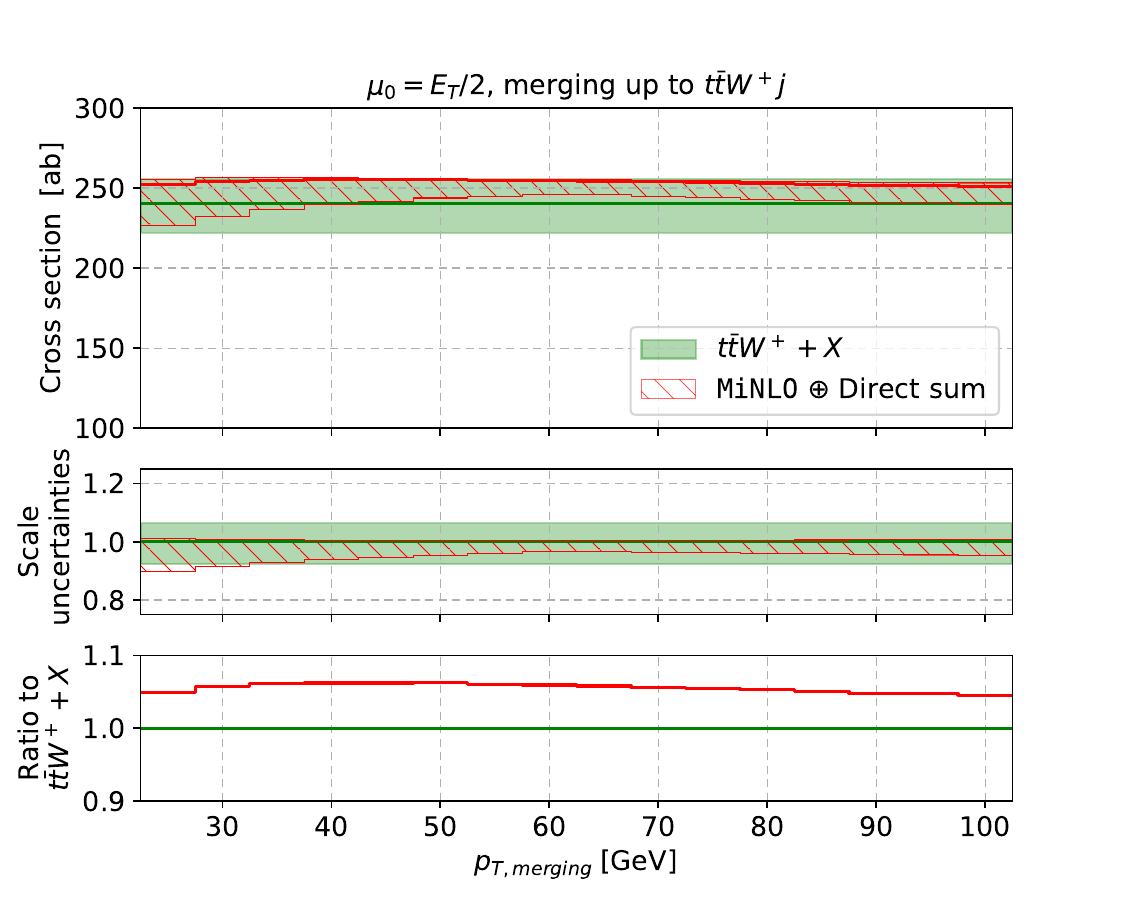}
        \includegraphics[width=0.48\linewidth]{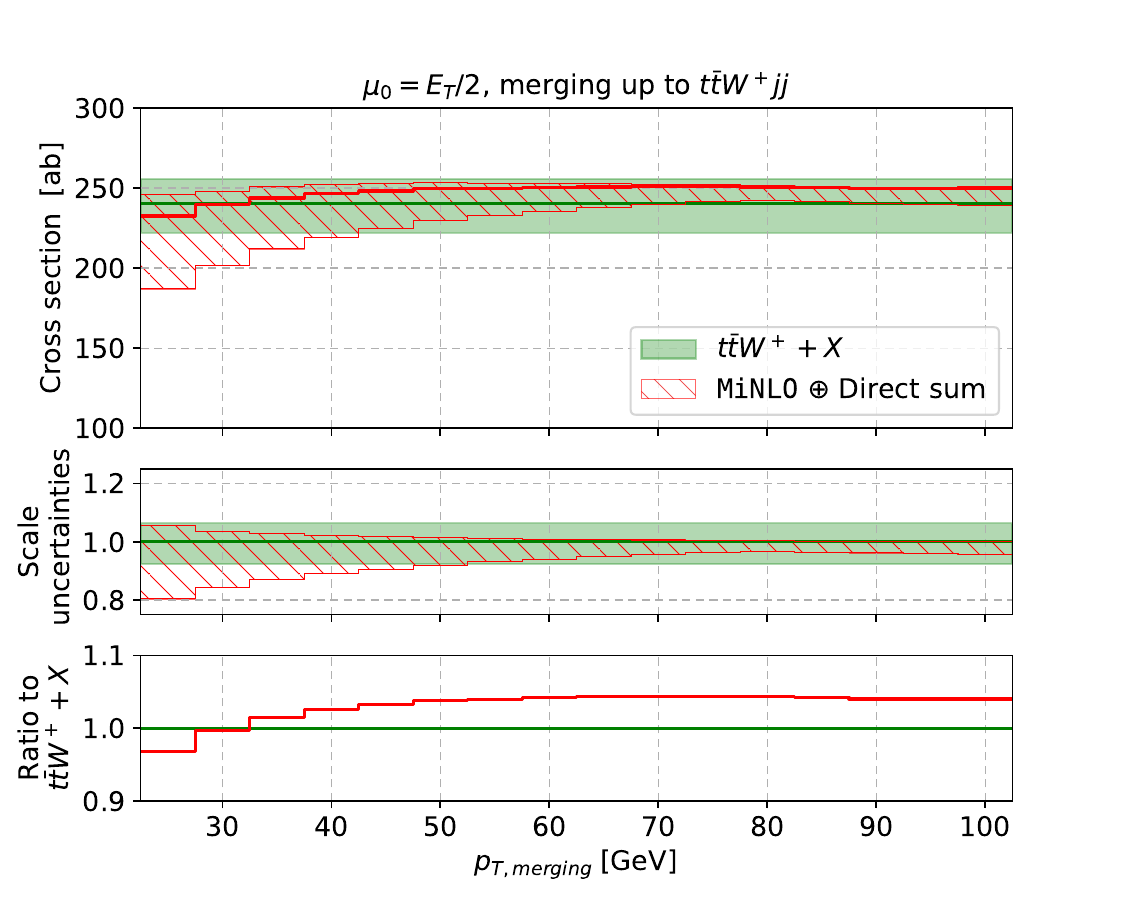}
        \caption{\textit{Multi-jet merged predictions, labeled as $\texttt{MiNLO} \, \oplus \,  \rm Direct \, sum$, as a function of several merging-scale parameters $p_{T,\, merging}$ for the full off-shell $pp \to t\bar{t}W^+ + \rm jets$ process at the LHC with $\sqrt{s} = 13\,\rm TeV$. The merging is performed including up to one (left) or two (right) resolved jet(s) and using the $\mu_0=E_T/2$ scale choice. The merged results are compared with the full off-shell NLO $pp \to t\bar{t}W^+ + X$ cross-section in the multilepton channel. The upper panels show the absolute predictions, the middle panels the corresponding scale uncertainties, and the bottom panels the ratio to the full off-shell $pp \to t\bar{t}W^+ + X$ prediction.}}
         \label{fig:merged_integrated}
\end{figure}

\begin{figure}[t!]
        \centering        
        \includegraphics[width=0.48\linewidth]{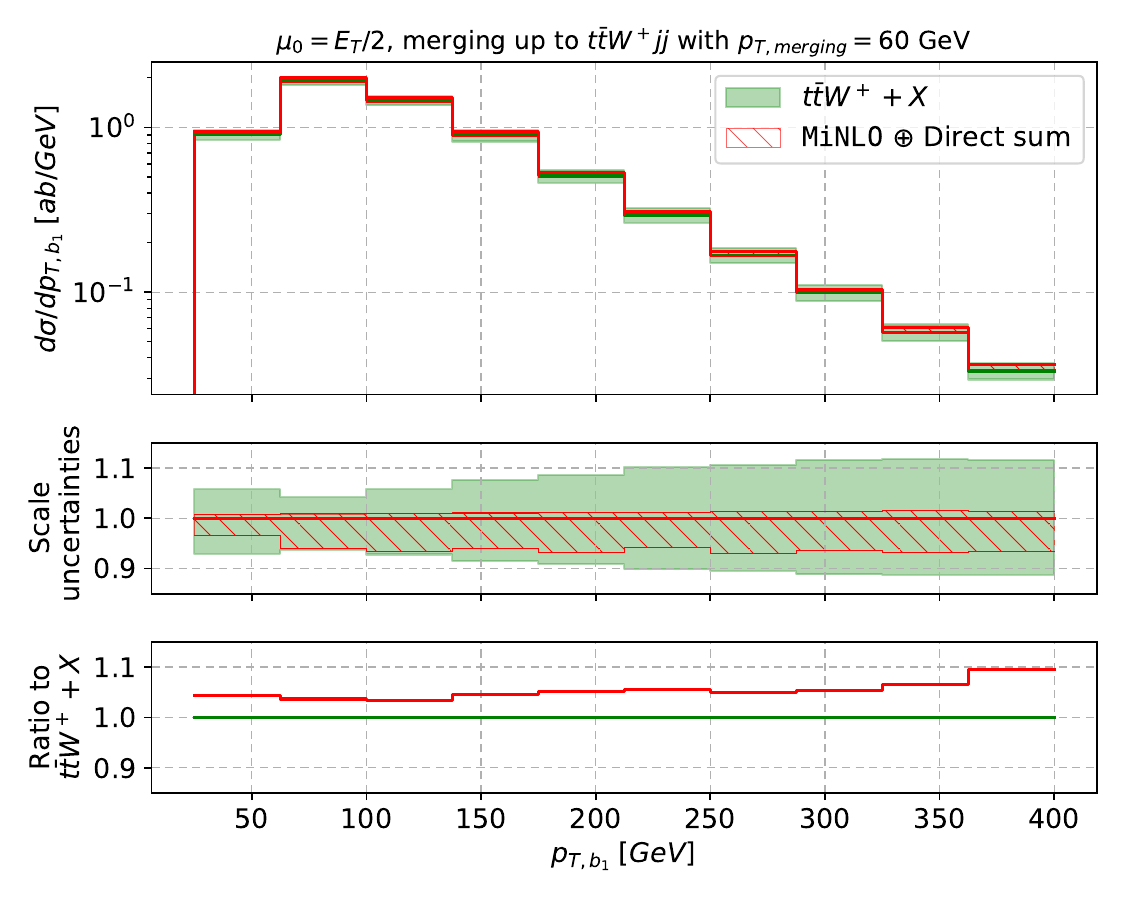}
        \includegraphics[width=0.48\linewidth]{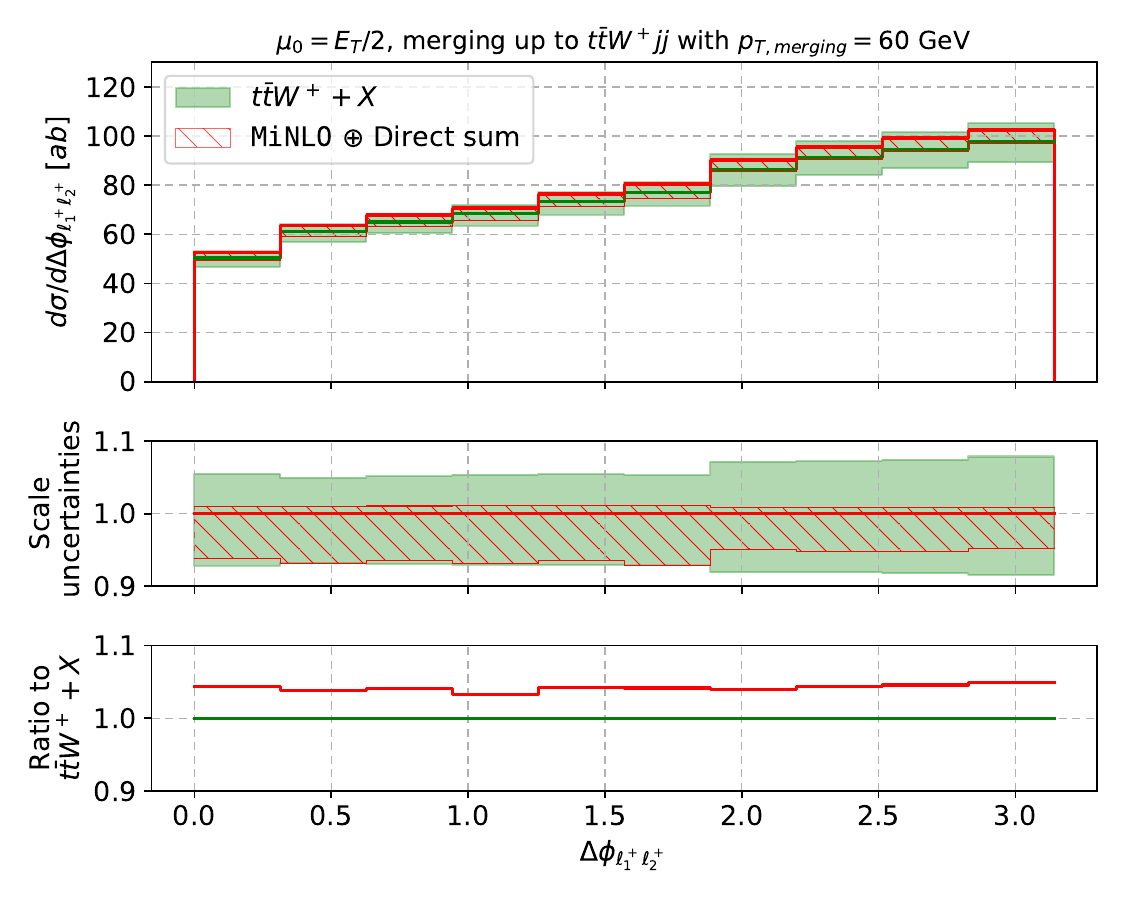}
        \caption{\textit{Differential merged predictions with up to 2 jets, labeled as $\texttt{MiNLO} \, \oplus \,  \rm Direct \, sum$, for the full off-shell $pp \to t\bar{t}W^+ + \rm jets$ process at the LHC with $\sqrt{s} = 13\,\rm TeV$. A merging-scale parameter equal to $p_{T,\,merging} = 60 \, \rm GeV$ has been used and the $p_{T,\,b_1}$ (left) and $\Delta \phi_{\ell_1^+ \ell_2^+}$ (right) observables are shown. The structure of the plots follows the one from Fig. \ref{fig:merged_integrated}.}}
         \label{fig:diff_merged_1}
\end{figure}

To improve the full off-shell predictions of the $pp \to e^+ \nu_e \mu^- \bar{\nu}_\mu \tau^+ \nu_\tau b\bar{b} + X$ process, it is important to combine samples with different jet multiplicities. Indeed, as previously mentioned, it has been shown that the additional jet activity is quite essential for a more realistic description of the $t\bar{t}W^+$ dynamics. The merging procedure employs a merging-scale parameter $p_{T,\,merging}$ to prevent double counting and is performed either up to one or two jets, where in the latter case, the LO $pp \to e^+ \nu_e \mu^- \bar{\nu}_\mu \tau^+ \nu_\tau b\bar{b}\,jj$ sample is used. In all samples, the \texttt{MiNLO} method is applied separately. This is in contrast to the original \texttt{MiNLO} merging \cite{Hamilton:2012rf}, where no merging scale is required. However, for the complicated full off-shell $pp \to e^+ \nu_e \mu^- \bar{\nu}_\mu \tau^+ \nu_\tau b\bar{b} + X$ process with fiducial phase-space cuts, the relevant resummation structure is not known. Therefore, the \texttt{MiNLO} merging procedure cannot be directly applied in its standard form, as it requires process-dependent modifications of the Sudakov form factors based on the corresponding resummation structure. The merged results, labeled as $\texttt{MiNLO} \oplus \rm Direct \, sum$, are demonstrated in Fig. \ref{fig:merged_integrated} and are compared with the fiducial full off-shell $pp \to t\bar{t}W^+ + X$ cross section.
We observe that for merging up to one jet (two jets), values of $p_{T,\,merging}$ above $40$ $(55)$ GeV yield merged predictions with reduced uncertainties compared to the $t\bar{t}W^+$ cross section, resulting in improved results. On the one hand, the difference between merged predictions and the $t\bar{t}W^+$ cross section remains relatively flat at the $5\%-7\%$ level for merging up to one jet. For merging up to two jets, on the other hand, the merged results are initially smaller at low $p_{T,\,merging}$ values but gradually increase, reaching deviations of at most $5\%$. It is also worth noting that when performing the merging up to 2 jets, the theory uncertainties are generally larger for small $p_{T,\, merging}$ values compared to the 1-jet merging case. This highlights the importance of including an NLO accurate $pp \to e^+ \nu_e \mu^- \bar{\nu}_\mu \tau^+ \nu_\tau b\bar{b}\,jj$ sample in the merging procedure, which is beyond the scope of this study. Interestingly, by doing so, the second leading jet will be described with NLO accuracy, thereby exceeding the formal accuracy of an NNLO calculation. After imposing appropriate values of the merging-scale parameters, the predictions are substantially improved also at the differential cross-section level, as shown in Fig. \ref{fig:diff_merged_1}, where the transverse momentum of the hardest $b$ jet and the azimuthal angle between the two positively charged leptons are illustrated. For the dimensionful observable, some shape differences are observed, particularly in the high-$p_T$ regime, but they do not exceed $10\%$. In contrast, the dimensionless observable shows only normalization differences.

\section{Conclusions} \label{sec:conclusions}
The comparison between standard NLO and \texttt{MiNLO} for the full off-shell $pp \to t\bar{t}W^+ j + X$ process in the multilepton channel shows that the two approaches yield consistent predictions within the uncertainties. While \texttt{MiNLO} does not significantly alter the results for well-behaved dynamical scale choices, it can substantially reduce the scale uncertainties associated with a fixed scale choice, both at the integrated and differential cross-section levels. Furthermore, the merging of samples with different jet multiplicities provides a more complete description of the full off-shell $pp \to t\bar{t}W^+ + X$ process, with reduced uncertainties after appropriate merging-scale parameters have been chosen.

\acknowledgments{This research was supported by the Deutsche Forschungsgemeinschaft (DFG) under the following grants: TRR 257 -  {\it P3H - Particle Physics Phenomenology after the Higgs Discovery} and GRK 2497 -  {\it The Physics of the Heaviest Particles at the LHC.}






\end{document}